\documentclass[runningheads]{llncs}
\usepackage[T1]{fontenc}
\usepackage{graphicx}
\usepackage{xspace}
\usepackage[hidelinks]{hyperref}
\usepackage{color}

\newcommand\Platform{Platform\xspace}

\begin{document}
\title{Reliably Reproducing Machine-Checked Proofs\\ with the Coq Platform}
\titlerunning{Reliably Reproducing Machine-Checked Proofs with the Coq Platform}
%
\author{Karl Palmskog\inst{1} 
 \and
Enrico Tassi\inst{2} 
 \and
Théo~Zimmermann\inst{3} 
}
\authorrunning{K. Palmskog, E. Tassi, and T. Zimmermann}
%
\institute{KTH Royal Institute of Technology, 100 44 Stockholm, Sweden\\
\email{palmskog@acm.org}\\
\and
Univ. C\^ote d'Azur, Inria, France\\
\email{enrico.tassi@inria.fr}
\and
Inria, Univ. Paris Cité, CNRS, IRIF, UMR 8243, F-75013 Paris, France\\
\email{theo@irif.fr}}
\maketitle              
\begin{abstract}
The Coq Platform is a continuously developed distribution of the Coq proof assistant together with commonly used libraries, plugins, and external tools useful in Coq-based formal verification projects. The Coq Platform enables reproducing and extending Coq artifacts in research, education, and industry, e.g., formalized mathematics and verified software systems. In this paper, we describe the background and motivation for the Platform, and outline its organization and development process. We also compare the Coq Platform to similar distributions and processes in the proof assistant community, such as for Isabelle and Lean, and in the wider open source software community.
\end{abstract}
\section{Introduction}
The Coq proof assistant~\cite{Coq815} provides a formal language to write datatypes, functions, and theorems, together with an environment for semi-interactive development of machine-checked proofs. Typical applications of Coq are in formalization of mathematics~\cite{Gonthier2013} and formal verification of software~\cite{Leroy2009,Appel2022}.

The Coq Platform (\emph{\Platform} for short) is a continuously developed distribution of Coq together with many commonly used Coq libraries and Coq plugins. The \Platform also provides several external tools, e.g., for proof search and automation~\cite{Czajka2018} and build management~\cite{Dune}. In addition to the latest version of Coq, each \Platform release supports several previous Coq versions. Thanks to Coq's code compatibility policy, this ensures that both new and legacy Coq-based artifacts in research, education, and industry can be reliably reproduced, and when necessary, upgraded and extended.

In this paper, we describe the background and motivation for the \Platform, which arguably traces its origin to a collection of standalone libraries (``Contribs'') from 1993. We then outline the \Platform's organization and development process. Through the adoption of Coq Enhancement Proposal (CEP)~52 in 2021~\cite{CEP52}, \Platform development has become an integrated part of the development process of Coq itself, with Coq core team members comprising half the \Platform team. The \Platform also provides release coordination for projects in the Coq ecosystem, both explicitly by asking maintainers for \Platform project releases, and implicitly by new \Platform releases prompting releases of non-\Platform projects.

Finally, we compare the \Platform to similar initiatives for reliably reproducing software artifacts, both in the proof assistant community and in the wider open source software community. Our description and comparisons are based on \Platform release 2022.01.0~\cite{Platform2021010}, which provides Coq versions 8.12 to 8.15, along with up to 50 curated and tested packages for each version.

\section{Coq Overview}

Coq is a proof assistant based on type theory, implemented mainly in the OCaml programming language. Coq was first publicly released in 1989, and is currently maintained as open source software on GitHub~\cite{CoqGitHub}. To end users, Coq provides an environment for purely functional programming, along with facilities for specification, reasoning, and proof checking.
Coq consists of several parts:
\begin{itemize}
\item Coq's surface-level language is the \emph{vernacular}, which is an extensible language of commands written as a sequence of \emph{sentences}. Commands can be queries into definitions, declarations of new definitions (e.g., of functions or datatypes) using the Gallina language, or \emph{proof tactics} to run.
\item Coq \emph{elaborates} user-written vernacular into a logical formalism, which is an extension of the Calculus of Inductive Constructions~\cite{Coquand1990}, a powerful foundational theory of dependent types.
\item After elaboration, Coq's \emph{kernel} certifies that a \emph{term} has a proposed \emph{type}. For example, a type may be the statement of a theorem, and the term its proof.
\end{itemize}

Both Coq's elaborator and kernel can take significant time to execute for some vernacular code, e.g., due to long-running custom tactics that build terms or heavy use of Coq's built-in proof search methods. Moreover, Coq tactic commands (proof scripts) are seldom written to explicate the intuition behind proofs.
Consequently, a user may not be convinced that a formal Coq proof (term) of a statement (type) exists unless it is reproducible locally. Hence, a typical mode of use of Coq is to check publicly available vernacular code related to specific applications in mathematics and computer science.

However, Coq code usually depends on many \emph{third-party libraries} in addition to Coq's own standard library (Stdlib). For example, Stdlib contains formalizations of natural numbers, integers, and real numbers, but formalizations of floating points and regular expressions are found in other libraries~\cite{Doczkal2018,Boldo2011}. Similarly, the built-in proof automation is often supplemented by tactics defined in \emph{plugins}.

The \Platform provides a way to install Coq and a curated selection of such libraries and plugins in a controlled way. Notably, the \Platform provides the CompCert C compiler~\cite{Leroy2009}, the Verified Software Toolchain~\cite{Appel2022}, and their dependencies, enabling formal verification of functional correctness of C programs down to the instruction set architecture (ISA) level in Coq.

\paragraph{Coq's code compatibility policy.}

Vernacular code that can be fully checked (compiled) on a Coq major version, e.g., 8.11, is not \emph{guaranteed} to work on the next major version, e.g., 8.12. However, since 2015, Coq development follows a \emph{compatibility policy}~\cite{compatpolicy,Zimmermann2019} which makes compilation errors for legacy code on new versions predictable in most cases. The policy essentially says that:
\begin{itemize}
\item It should be \emph{possible} to be compatible with two successive major versions.
\item Features should be deprecated in one major version before removal.
\item Developers should provide an estimate of the required effort to fix a project with respect to a given change in Coq.
\item Breaking changes should be clearly documented in the public release notes, along with recommendations on how to fix a project if it breaks.
\end{itemize}
This means, for example, that a Coq artifact compatible with Coq 8.13 can with reasonable effort be made compatible first with 8.14 and then 8.15, using detailed instructions from deprecation warnings and release notes. 

\section{\Platform Usage}

Since 2021, the \Platform is the recommended way to install Coq on Coq's website~\cite{CoqWebsite}, since the \Platform can accommodate a large range of use cases across variants of Windows, macOS, and Linux: 
\begin{itemize}
\item \emph{Binary installers} are the simplest and fastest way for a new user to get a working installation of Coq together with many third-party packages. However, its main limitation is that it is not customizable, i.e., users cannot change the set of packages available to them.
\item \emph{Interactive scripts} provide a cross-platform solution to get Coq and third-party packages installed with the opam package manager~\cite{opam}, by handling the installation of any required system dependencies and running the appropriate opam commands. After basic installation, users can customize the set of available packages (and their versions) by using opam directly.
\end{itemize}

Hence, as long as people using the scripts are not editing the resulting set of packages, the \Platform provides a reliable cross-platform solution for ensuring that people running different operating systems get the same set of Coq package versions. This can simplify the life of professors teaching Coq to their students, but also of scientific reviewers trying to reproduce a Coq artifact. However, compatibility between packages is only assured when using \Platform release package versions. Customizing package versions may lead (back) to ``dependency hell''~\cite{Appel2022}.

\section{\Platform History and Goal}

\subsection{Coq Contribs and the Coq opam archive}

Coq Contribs (CCs), which are a collection of standalone libraries and plugins for Coq,
have been around since 1993~\cite{Zimmermann2019}. CCs followed a model where
the author of a Coq artifact hands over all maintenance to Coq developers.

CCs were always available for public download, but the repository hosting them was private
in the sense that no contributor other than the Coq developers had
write access (thus, excluding the original author). Consequently, active projects
lived outside of this walled garden, possibly leading to split evolution.

Around 2013, Thomas Braibant came up with the idea of using OCaml's package
manager, opam~\cite{opam}, for Coq packages. This idea was implemented one year later by
Guillaume Claret and the second author of this paper in the form of the
Coq opam archive on GitHub~\cite{CoqOpamArchive} and the Coq Package Index~\cite{CoqPackageIndex}, which together have supplanted the (now stale) CCs.
Officially adopting a package manager greatly simplified the distribution of
libraries from Coq users to other Coq users, and currently the Coq opam archive hosts
thousands of packages, including most of the current \Platform packages. 
The remaining \Platform packages are hosted in the general OCaml opam archive.

\subsection{Coq's Windows installer}

Historically, Coq has primarily been developed for Unix-like operating systems,
and its implementation language OCaml has had limited Windows support.
As a result, it was always necessary to provide a pre-built Windows installer.

In 2017, Michael Soegtrop 
started to include third-party Coq packages
into the Windows installer to encourage evaluation and eventually adoption of Coq
in industry. He also started to distribute the sources that were used
to build the installer for reproducibility (and licensing) concerns.
This augmented installer received positive feedback, in particular from non-academic
users. In fact, many users would not consider using any library not
part of the installer---both due to installation difficulty
and the risk of unavailability for new Coq versions.

In 2019, following various discussions and experimentations,
Soegtrop announced the
\Platform project at a meeting of Coq developers in Nantes, and published the first revision of its
charter~\cite{Charter}. This charter was inspired by previous work on the Windows installer,
but extended its scope to also support macOS and Linux and include additional packages.

Up until the release of Coq 8.12 in summer 2020, the Coq release process included
the task of building the Windows installer and a macOS dmg archive. The effort of
this task had grown considerably over time, and the skills and motivation
needed in order to fix related problems were not abundant in the
Coq core team (from which the Coq release managers are chosen).
In addition, the fact that the Windows installer shipped with several external packages created a synchronization problem, since Coq developers had to select compatible versions of some packages \emph{before} the corresponding Coq release.

At the time when Coq 8.13 was supposed to be released, in January 2021, the
Coq scripts to build binary installers were not functional any more, while
the \Platform installers were functional. As a result, 8.13
was the first release of Coq where binary packages were built using the \Platform. 
%
These events led to the publication of CEP~52~\cite{CEP52} by the second author,
which documents how the release of Coq and the release of the \Platform is
currently organized; this is described in detail in Section~\ref{sec:platformreleasecycle}.

\subsection{\Platform goal and problems addressed}

The current goal of the \Platform is to provide a distribution for developing and teaching with Coq that is \emph{operating-system independent}, \emph{dependable}, \emph{easy to install}, and \emph{comprehensive}. As part of the process of achieving this goal, we believe the \Platform can partially address a number of long-standing problems that we have experienced in the Coq community and ecosystem:
\begin{description}
\item[Release coordination.] Coq libraries may not provide releases for a new Coq version. By being included in the \Platform, project maintainers are incentivized to provide releases and can more easily collaborate with Coq developers and users to support new Coq versions.
\item[Compatibility.] Coq libraries and plugins may not provide mutually compatible releases. In the \Platform, packages are continually tested for compatibility, and \Platform releases do not include mutually exclusive packages.
\item[Build automation.] Many Coq projects do not use best practices for Coq building and testing, and may not work across different operating systems. The \Platform can help disseminate knowledge of such practices, and tests its packages on several operating systems.
\item[Reproducibility of research artifacts.] Coq research artifacts have historically been distributed as opaque vernacular file collections (e.g., tarballs) without explicit Coq version compatibility. By advising authors to target and document a specific \Platform release, scientific venues can ensure Coq artifact reproducibility.
\item[Upgrade paths between Coq versions.] Coq users may hold off upgrading to a new Coq version because one of the libraries they depend on is not available. The \Platform aims to provide an upgrade path from one Coq version to another with a consistent collection of packages that should only increase over time. To ensure this, authors of packages that join the \Platform must agree to a form of social contract, which, e.g., entails making timely releases and collaborating with the \Platform maintainers to solve user issues.
\end{description}

\section{\Platform Development and Organization}

\subsection{Platform development and release process}
\label{sec:platformreleasecycle}

Ever since the adoption of CEP~52, the release processes of Coq and the \Platform (and its
binary installers) follow approximate 6-month cycles and are handled separately, by different, possibly overlapping, teams. This simplifies the tasks of Coq release managers, who can focus on the piece of software they know best.

The work of the Coq Platform team begins when the Coq core team publishes a new major Coq version release candidate, on top of which the \Platform can then be built. When the new stable Coq version is finalized a few weeks later, it is not announced to the user community; this happens only when the Coq version becomes included in a \Platform release.

In order to produce a new \Platform release, maintainers of \Platform packages need to be involved. As soon as a Coq release candidate is published, it is made available in a development-only opam
repository, and corresponding Docker images are built to facilitate compatibility testing in continuous integration workflows.
The maintainers of
packages that are part of the \Platform are informed by the \Platform team that a new release is happening and that their participation is needed to determine the version  of their package that will be included. Based on the tests performed on the
\Platform (see Section~\ref{sec:platformci}), they may be informed that a certain package version or commit is already compatible, or that there are no known compatible version of their packages.
In both cases, they should decide which version they would like to have included, and make a new package release if needed.
So far, for three \Platform releases, the community of package maintainers has been responsive
to this call, accepting the proposed version or providing a new working version
in a matter of weeks, although, for a few large and complex packages, the \Platform
team had to assist the developers.

\subsection{Platform versioning scheme and organization}

During the process of adopting CEP 52, it was decided that the \Platform would use a calendar-based versioning scheme, independent of the Coq versions it includes, to reflect that the \Platform is a distribution, not only of Coq, but also of many other packages that have their own release cycle.
Since September 2021, this has allowed an additional (technical and organizational) change in the way the \Platform is maintained and published. Instead of being tied to a specific Coq version, \Platform releases now include several ``package picks'', and the user can decide which one to install by selecting the appropriate binary installer, or the appropriate option in the interactive scripts.

This change serves several purposes. From the user point of view, old package picks are still available in the next versions of the \Platform, and thus users can always rely on the latest version. The interactive scripts make it possible to install several picks in parallel, and thus, this provides a smooth upgrade experience from one pick (and one Coq version) to the next. From the \Platform maintainers' point of view, this allows maintaining several Coq versions and package picks as part of a single branch, and thus to factorize any improvements to the infrastructure, but also any fixes that are independent of the Coq version (for instance, when a new Cygwin version or Ubuntu Linux version introduces changes that break the compilation of Coq dependencies, such as OCaml).

\subsection{\Platform continuous integration and delivery}
\label{sec:platformci}

The \Platform repository uses GitHub Actions for continuous
integration and delivery (CI/CD). Initially, CI/CD only built binary installers for
Windows. Then, it was then extended to build the MacOS dmg and finally a Snap package
for Linux. CI/CD is also used to test the interactive installers.
The \Platform finally includes a ``smoke test kit'', a test suite that runs after
exercising the binary installers. The test suite checks that Coq and all included packages
can be actually loaded and used, which can catch problems in the installers themselves.


CI was also greatly simplified when the same branch and scripts were able to
support multiple package picks. The CI configuration files are thus factorized
as well, and rely on the common ``matrix'' feature to test multiple picks.

Finally, users have used the CI/CD setup to build custom \Platform installers,
typically to override package versions and include extra packages for teaching
purposes. By forking the repository, performing changes to the package pick
definitions, and then letting CI run (in their fork or in a draft pull request
on the main repository), they can produce binary installers for all the major
operating systems, without needing to have these systems at hand.

\section{\Platform Role in the Coq Ecosystem}

Historically, CCs served several roles at once: distribution of third-party developments to the users, long-term maintenance, and testing. In the Coq ecosystem as of early 2022, these roles have been split into three parts:

\begin{description}
\item[Coq Package Index.] The main channel for distributing third-party Coq packages is using the opam-based Coq Package Index. In contrast to CCs, the source code for a package in the index is under the full control of its authors.
\item[Coq-community.] Useful packages may stop being maintained by their original authors, e.g., when produced for a research paper or as part of a PhD thesis. When this happens, interested users can carry on maintenance by adopting the package in the Coq-community organization on GitHub~\cite{coqcommunity}. As of 2022, Coq-community hosts around 60 projects maintained by around 30 people; around 20 projects are former CCs.
\item[Coq's CI test suite.] Coq developers test their changes to Coq against a test suite consisting of many (actively maintained) external projects. When they decide to merge a change that breaks a project in this suite, they write compatibility fixes for this project and submit it to the project's maintainers. This means that, in practice, Coq developers still participate to the long-term maintenance of some important Coq packages. However, they are only responsible for producing compatibility fixes, but the roadmap and evolution of the packages and their release schedule still remain under the control of their original authors (or current maintainers).
\end{description}

In this context, the role of the \Platform is not only to provide a convenient way to install Coq along with packages found on the Coq Package Index, but also to curate generally useful packages whose maintainers agree to the \Platform's social contract. The coordination signals transmitted by \Platform releases also help the Coq ecosystem catch up to new versions more quickly than previously, when few packages were compatible with a new Coq version following its release~\cite{ExportCoq}.

As Coq projects evolve, they may become incompatible with other (dependent) projects, e.g., due to breaking changes that are difficult to accommodate. Since maintainers of such projects have committed to long-term \Platform inclusion, they have an incentive to find solutions to these issues together with \Platform maintainers, e.g., through the introduction of new packages that facilitates slow-paced user migration away from a legacy package.

\section{Comparisons}

\subsection{Ecosystem-specific platforms and distributions}

The Coq Platform is comparable to many similar software ``platforms'' and distributions which it takes inspiration from (e.g., the Haskell Platform, the Scala Platform, and TeXLive). These distributions are all centered around the concept of making it easier to access a collection of third-party packages and providing a ``batteries included'' experience to beginners.

However, while some distributions such as TeXLive have been largely successful, some others have not: the Haskell Platform was deprecated in 2022. Many technical and social factors can enter into consideration and result in eventual failure or success, and the reasons why the Haskell Platform was deprecated are not documented. Nevertheless, the Haskell ecosystem has another similar maintained solution for providing packages with compatible versions: Stackage~\cite{Stackage}. Similarly to the Coq Platform, Stackage is based on a social contract~\cite{StackageMaintainers}.

\subsection{Linux distributions}

Linux distributions provide consistent sets of packages for important software suites, including proof assistants such as Coq, and sometimes also libraries. However, in rapidly evolving ecosystems, it quickly becomes impossible to provide a large set of compatible libraries without coordination with their maintainers. When coordination mechanisms are put in place (such as with the Coq Platform, or with Stackage), this can provide ways for distribution maintainers to provide a larger set of packages, by relying on the documented sets of compatible versions. This is why Linux distributions virtually always provide a package for TeXLive. Similarly, we expect that (if there was enough interest for Coq libraries), distribution maintainers could provide packages matching the content of the \Platform. We aim to provide documentation to facilitate this endeavor.

\subsection{Isabelle and the Archive of Formal Proofs}

The Isabelle generic proof assistant, and in particular its Isabelle/HOL instantiation, are the basis for the Archive of Formal Proofs (AFP)~\cite{IsaAFP}. Entries to the AFP are submitted by authors and then reviewed both on content and using technical criteria. Updates to AFP entries are primarily done by Isabelle developers when code breaks due to Isabelle evolution, rather than by the authors of the entries. While some Coq developers participate in \Platform maintenance, they normally only update projects that are in Coq's CI test suite; not all \Platform projects are in this test suite, and it contains many non-\Platform projects.

The focus in the \Platform is on including generally useful libraries, plugins, and tools---not novel research artifacts that are a focus of the AFP. Coq research artifacts can instead be submitted to the Coq opam archive. However, the Coq opam archive maintainers only perform minimal reviewing and updating of submitted packages. Each \Platform release includes several Coq versions, while the AFP works with a single Isabelle version at any given time.

\subsection{Lean and mathlib}

The Lean prover~\cite{Lean3,Lean4} has similar foundations to Coq, but its development and ecosystem is differently organized in several important ways. In contrast to Coq, where developers provide upgrade paths between versions, Lean 3 and Lean 4 are to a large extent incompatible; support for porting code is nevertheless provided by tools developed in the Lean user community~\cite{Mathport}.

Lean has a single significant large library, mathlib~\cite{Mathlib}, that contains an extensive collection of formalized mathematics and is actively developed in a single repository on GitHub~\cite{MathlibGitHub}. The monorepository approach used for mathlib comes with several advantages, not least of which are avoiding version management. Coq projects must continually decide whether they should do a release compatible with a new Coq version, while mathlib is continually compatible with the latest release of Lean 3. The switch of mathlib from Lean 3 to Lean 4 is expected to happen atomically, rather than using versioning as for Coq libraries.

Thanks to the cross-platform binary format for compiled code used by Lean, mathlib can be easily distributed both in source and binary form, which generally leads to a faster setup than for comparable Coq libraries which are compiled from source. However, we believe the operating-system-specific binary distributions of the \Platform have comparable ease of use.

\section{Conclusion}

In this paper, we presented the history, current state, and organization of the Coq Platform,
the official distribution of the Coq proof assistant, and its aim to improve
Coq-based artifact reproducibility and solve other coordination problems in the Coq
ecosystem.

In recent years, venues such as POPL, IJCAR, FLOC and ETAPS have seen an
increasing number of submissions supported by Coq artifacts.
In specialized conferences such as CPP and ITP, nearly all submissions come
with a proof assistant artifact as their main contribution, many of
which use Coq.
We believe that current standards for evaluating such artifacts
are too low even at CPP and ITP. In particular, there is no strict requirement
for the artifact to be reproduced by the reviewers. In fact, the program committee
chairs are usually more than happy if one reviewer (out of typically three) manages
to do it, since recreating an environment where the artifact can actually
be inspected can be extremely difficult and time consuming.

The Platform will hopefully make the task of evaluating Coq artifacts easier and quicker,
not only in specialized venues, but also in broader ones where program committee members may have less Coq experience.

%
%
%



\subsubsection{Acknowledgements}
We are grateful to Michael Soegtrop, all members of the Coq Team, and the developers of all \Platform packages. We also thank the Lean community for feedback.

%
%
%
\bibliographystyle{splncs04}
\bibliography{bib}

\end{document}